# Linear Canonical Transformations in Relativistic Quantum Physics

**Ravo Tokiniaina Ranaivoson[1], Raoelina Andriambololona[2], Hanitriarivo Rakotoson[3], Roland Raboanary[4]**

*tokhiniaina@gmail.com[1], tokiniainaravor13@gmail.com[1], raoelina.andriambololona@gmail.com[2], jacquelineraoelina@hotmail.com[2], raoelinasp@yahoo.fr[2], infotsara@gmail.com[3], r_raboanary@yahoo.fr[4]*

[1,2,3]*Information Technology and Theoretical Physics Department*
*Institut National des Sciences et Techniques Nucléaires (INSTN- Madagascar)*
BP 3907 Antananarivo 101, Madagascar, *instn@moov.mg*

[1, 2,3]TWAS *Madagascar Chapter, Malagasy Academy*
BP 4279 Antananarivo 101, Madagascar

[3,4]*Faculty of Sciences – University of Antananarivo*
BP 566 Antananarivo 101, Madagascar

***Abstract:*** Linear Canonical Transformations (LCTs) are known in signal processing and optics as the generalization of certain useful integral transforms. In quantum theory, they can be identified as the linear transformations which keep invariant the canonical commutation relations characterizing the coordinates and momenta operators. In this work, the possibility of considering LCTs to be the elements of a symmetry group for relativistic quantum physics is studied using the principle of covariance. It is established that Lorentz transformations and multidimensional Fourier transforms are particular cases of LCTs and some of the main symmetry groups currently considered in relativistic theories can be obtained from the contractions of LCTs groups. It is also shown that a link can be established between a spinorial representation of LCTs and some properties of elementary fermions. This link leads to a classification which suggests the existence of sterile neutrinos and the possibility of describing a generation of fermions with a single field. Some possible applications of the obtained results are discussed. These results may, in particular, help in the establishment of a unified theory of fundamental interactions. Intuitively, LCTs correspond to linear combinations of energy-momentum and spacetime compatible with the principle of covariance.

## 1-Introduction

Linear Canonical Transformations (LCTs) are studied and used in several areas like signal processing, optics, and quantum physics. In the fields of signal processing and optics, they are known to be the generalization of some useful integrals transforms such as Fourier and Fractional Fourier Transformations. It can be established (as shown in the appendix) that these integral transforms are equivalent, in quantum theory, to linear transformations of coordinates and momenta operators which keep invariant the canonical commutations relations. This equivalence makes easy the multidimensional generalization. Noting that the canonical commutation relations can be considered as fundamental relations in relativistic quantum physics, our main purpose in the present work is to study the possibility of considering the group of the LCTs to be a symmetry group using the principle of covariance.

It is known that the concept of symmetry plays an important role in physics and most major theories in physics have symmetry groups with associated fundamental invariants and covariants physical laws. It has been discussed for instance in the references [1-3]. For classical mechanics, the main symmetry group is the Galilei's one and some of the main invariants and covariants are mass, time, Newton's laws and the equations of analytical mechanics. For Special Relativity and relativistic Quantum Field Theory [4], the main symmetry group is the Lorentz group or the Poincaré group (if spacetime translations are included) and some of the main associated invariants and covariants are the speed of light, the pseudo-distance in spacetime or more generally any Minkowskian inner product and all the fundamental laws of related relativistic theories. For the de Sitter invariant Special Relativity and related theories, the symmetry group is the de Sitter group [5-9]. These facts can be considered as related to the principle of covariance which states that the laws of physics should have the same form in all admissible frames of reference.

In this work, the possibility of considering the group of Linear Canonical Transformations (LCTs) to be an adequate symmetry group for relativistic quantum physics is studied. Outside of the context of relativistic quantum physics, many works have already been done regarding linear canonical transformations and their applications (see for instance refs. [10-16]). In the section 2, the context of the approach is explained by defining LCTs in the framework of relativistic quantum physics. In the section 3, the identification of simplest particles quantum states that can be associated naturally with the LCT symmetry is performed. These states can be considered as generalizations of the well-known ground state of a harmonic oscillator. They share similarities with what are called coherent states and squeezed states in the literature [17-20]. In the section 4, it is shown that a link can be established between some properties of elementary fermions and a spinorial representation of LCTs. The purpose is to highlight, in a first study, the possibility of the obtaining of this kind of link and its importance for fermions field theory. This link permits to establish a classification of a generation of fermions according to the values of their charges. This classification suggests the existence of sterile neutrinos and the possibility of describing a generation of fermions with a single field. It is also remarked, in this section, that there is a similarity between the Bogolioubov Transformation [21-22] and the LCTs considered through this work. In the section 5, the obtaining of the indefinite orthogonal group $O(N_+, N_-)$ from the contraction of an LCT group is studied. The particular examples of a quandridimesional and a pentadimensional theories are considered and it is shown that the Lorentz group, the Poincare group and the de Sitter group can be obtained from contractions of an LCT group when quantum and gravitational effects are neglected. The main results obtained through this work are listed and discussed briefly with some of their possible applications and perspectives through the sections 6 and 7. These results may in particular help in the establishment of a unified theory of fundamental interactions which include gravity.

In the appendix, it is shown that the LCTs are equivalent to, and multidimensional generalization of, the well-known integral transforms which generalizes the Fourier and fractional Fourier transforms. The notation system used in this work is inspired from [23]. Boldface type corresponds to quantum operators. Einstein summation convention is used.

## 2- LCTs in the framework of relativistic quantum physics

In relativistic quantum physics, the spacetime coordinates operators $\boldsymbol{x}_\mu$ and energy -momenta operators $\boldsymbol{p}_\mu$ can be characterized by the canonical commutation relations. In natural unit system commonly used in many relativistic quantum theories (in which one takes for the reduced Planck constant $\hbar = 1$ and the speed of light $c = 1$), these canonical commutation relations are

$$\begin{cases} [\boldsymbol{p}_\mu, \boldsymbol{x}_\nu] = \boldsymbol{p}_\mu \boldsymbol{x}_\nu - \boldsymbol{x}_\nu \boldsymbol{p}_\mu = i\eta_{\mu\nu} \\ [\boldsymbol{p}_\mu, \boldsymbol{p}_\nu] = \boldsymbol{p}_\mu \boldsymbol{p}_\nu - \boldsymbol{p}_\nu \boldsymbol{p}_\mu = 0 \\ [\boldsymbol{x}_\mu, \boldsymbol{x}_\nu] = \boldsymbol{x}_\mu \boldsymbol{x}_\nu - \boldsymbol{x}_\nu \boldsymbol{x}_\mu = 0 \end{cases} \tag{1}$$

$\eta_{\mu\nu}$ being the covariant components of the symmetrical bilinear form defining the inner product i.e. the metric tensor. The relations (1) can be considered for a general theory corresponding to a pseudo-Euclidian space with signature $(N_+, N_-)$ and dimension $N = N_+ + N_-$. For the space used in the current formulation of special relativity and relativistic quantum field theory [4], i.e. the Minkowski space, the signature is $(1, 3)$ which means $N_+ = 1, N_- = 3$, $\eta_{00} = 1$, $\eta_{11} = \eta_{22} = \eta_{33} = -1$ and $\eta_{\mu\nu} = 0 \;\; if \mu \neq \nu$.

The canonical commutation relations (1) can be considered as fundamental relations for relativistic quantum physics defining spacetime coordinates operators $\boldsymbol{x}_\mu$ and energy-momenta operators $\boldsymbol{p}_\mu$ for an observer corresponding to a given frame of reference. For a second observer, using another frame of reference, the spacetime coordinates and energy-momenta operators, denoted respectively $\boldsymbol{x}'_\mu$ and $\boldsymbol{p}'_\mu$, are also required to satisfy these canonical commutation relations in agreement with the principle of covariance. Explicitly, we must have

$$\begin{cases} [\boldsymbol{p}'_\mu, \boldsymbol{x}'_\nu] = [\boldsymbol{p}_\mu, \boldsymbol{x}_\nu] = i\eta_{\mu\nu} \\ [\boldsymbol{p}'_\mu, \boldsymbol{p}'_\nu] = [\boldsymbol{p}_\mu, \boldsymbol{p}_\nu] = 0 \\ [\boldsymbol{x}'_\mu, \boldsymbol{x}'_\nu] = [\boldsymbol{x}_\mu, \boldsymbol{x}_\nu] = 0 \end{cases} \tag{2}$$

It is expected that a symmetry group corresponding to relativistic quantum physics is a group of transformations satisfying the relations (2) i.e. which keep invariant the canonical commutation relations. These transformations are the canonical transformations in relativistic quantum physics. The simplest kind of canonical transformations are linear ones (LCTs)

$$\begin{cases} \boldsymbol{p}'_\mu = \mathbb{a}_\mu^\nu \boldsymbol{p}_\nu + \mathbb{b}_\mu^\nu \boldsymbol{x}_\nu \\ \boldsymbol{x}'_\mu = \mathbb{c}_\mu^\nu \boldsymbol{p}_\nu + \mathbb{d}_\mu^\nu \boldsymbol{x}_\nu \end{cases} \tag{3}$$

A combination of the relations (2) and (3) permits to deduce that the constraints fulfilled by the parameters $\mathbb{a}_\mu^\nu, \mathbb{b}_\mu^\nu, \mathbb{c}_\mu^\nu, \mathbb{d}_\mu^\nu$ are

$$\begin{cases} \mathbb{a}_\mu^\rho \eta_{\rho\lambda} \mathbb{d}_\nu^\lambda - \mathbb{b}_\mu^\rho \eta_{\rho\lambda} \mathbb{c}_\nu^\lambda = \eta_{\mu\nu} \\ \mathbb{a}_\mu^\rho \eta_{\rho\lambda} \mathbb{b}_\nu^\lambda - \mathbb{b}_\mu^\rho \eta_{\rho\lambda} \mathbb{a}_\nu^\lambda = 0 \\ \mathbb{c}_\mu^\rho \eta_{\rho\lambda} \mathbb{d}_\nu^\lambda - \mathbb{d}_\mu^\rho \eta_{\rho\lambda} \mathbb{c}_\nu^\lambda = 0 \end{cases} \tag{4}$$

If the $N \times N$ matrices $\mathbb{a}, \mathbb{b}, \mathbb{c}, \mathbb{d}$ and $\eta$ corresponding to the coefficients $\mathbb{a}_\mu^\nu$, $\mathbb{b}_\mu^\nu$, $\mathbb{c}_\mu^\nu$, $\mathbb{d}_\mu^\nu$ and $\eta_{\mu\nu}$ are introduced, the relations in (4) are equivalent to the following matrix relations

$$\begin{cases} \mathbb{a}^T\eta\mathbb{d} - \mathbb{b}^T\eta\mathbb{c} = \eta \\ \mathbb{a}^T\eta\mathbb{b} - \mathbb{b}^T\eta\mathbb{a} = 0 \\ \mathbb{c}^T\eta\mathbb{d} - \mathbb{d}^T\eta\mathbb{c} = 0 \end{cases} \Leftrightarrow \begin{pmatrix} \mathbb{a} & \mathbb{c} \\ \mathbb{b} & \mathbb{d} \end{pmatrix}^T \begin{pmatrix} 0 & \eta \\ -\eta & 0 \end{pmatrix} \begin{pmatrix} \mathbb{a} & \mathbb{c} \\ \mathbb{b} & \mathbb{d} \end{pmatrix} = \begin{pmatrix} 0 & \eta \\ -\eta & 0 \end{pmatrix} \tag{5}$$

The relation (5) means that the matrix $\begin{pmatrix} \mathbb{a} & \mathbb{c} \\ \mathbb{b} & \mathbb{d} \end{pmatrix}$ belongs to the symplectic group $Sp(2N_+, 2N_-)$ [24-26]. This fact permits to have the identification of the LCT group, that we will denote $\mathbb{T}$, with $Sp(2N_+, 2N_-)$. For the case of the signature $(1,3)$, the group is $Sp(2,6)$.

**Remarks**

- The relations (1) and (3), considered in the framework of relativistic quantum physics, rise the problem of the existence of a time operator. This problem was already tackled by various authors as can be seen in the references [27-33]. The results established in the present work also show that the introduction of this operator could lead to interesting consequences.
- In the case $\mathbb{a} = \mathbb{d}$ and $\mathbb{c} = -\mathbb{b}$, the relations (1) and (5) defining the LCTs are reduced to

$$\begin{cases} \boldsymbol{p}'_\mu = \mathbb{a}_\mu^\nu \boldsymbol{p}_\nu + \mathbb{b}_\mu^\nu \boldsymbol{x}_\nu \\ \boldsymbol{x}'_\mu = -\mathbb{b}_\mu^\nu \boldsymbol{p}_\nu + \mathbb{a}_\mu^\nu \boldsymbol{x}_\nu \\ \mathbb{a}^T\eta\mathbb{a} + \mathbb{b}^T\eta\mathbb{b} = \eta \\ \mathbb{a}^T\eta\mathbb{b} - \mathbb{b}^T\eta\mathbb{a} = 0 \end{cases} \tag{6}$$

Two remarkable particular cases of the LCTs fulfilling (6) can be identified

**i)** For the case $\mathbb{b} = 0$ the LCTs (6) is reduced to

$$\begin{cases} \boldsymbol{p}'_\mu = \mathbb{a}_\mu^\nu \boldsymbol{p}_\nu \\ \boldsymbol{x}'_\mu = \mathbb{a}_\mu^\nu \boldsymbol{x}_\nu \\ \mathbb{a}^T\eta\mathbb{a} = \eta \end{cases} \tag{7}$$

The relation $\mathbb{a}^T\eta\mathbb{a} = \eta$ means that the $N \times N$ matrix $\mathbb{a}$ belongs to $O(N_+, N_-)$. For the signature $(1,3)$, this group is the Lorentz group $O(1,3)$ i.e. Lorentz transformations can be considered as particular cases of LCTs.

**ii)** For the case $\mathbb{a} = 0$, the transformation (6) is reduced to

$$\begin{cases} \boldsymbol{p}'_\mu = \mathbb{b}_\mu^\nu \boldsymbol{x}_\nu \\ \boldsymbol{x}'_\mu = -\mathbb{b}_\mu^\nu \boldsymbol{p}_\nu \\ \mathbb{b}^T\eta\mathbb{b} = \eta \end{cases} \tag{8}$$

The transformations (8), which transform spacetime coordinates operators into energy-momentum operators and inversely, are the multidimensional generalization of Fourier transforms. This fact is more explicit for the case $\mathbb{b} = I_N$ (the $N \times N$ identity matrix). This claim is justified in the appendix section: the LCTs (3) can be considered as the multidimensional generalizations of operator transformations which are equivalent to the integral transforms known as the LCTs in signal processing and optics.

These integral transforms give as particular cases the Fourier and fractional Fourier transforms. If an LCT group is considered as a symmetry group relating different frame of references, in the framework of relativistic quantum physics, the associated integrals transforms can be considered as corresponding to the laws of transformations of the wavefunctions associated to each frame of references.

## 3- Simplest particle states associated with the LCT symmetry

### 3.1 Mean values and statistical variance-covariances of momenta and coordinate operators associated to a particle state

Our goal in this section is to identify the simplest states of a particle that can be associated naturally with the LCT symmetry i.e. corresponding to wavefunctions which keep a "covariant form" under the action of an LCT. The introduction of the mean values and statistical variance-covariances of energy-momenta and spacetime coordinates operators is useful for this purpose.

Let us consider any quantum state $|\varphi\rangle$ of a particle. The following parameters can be defined

- the mean values of the operators $\boldsymbol{p}_{\mu}$ and $\boldsymbol{x}_{\boldsymbol{\mu}}$ associated with the state $|\varphi\rangle$, denoted respectively $\langle p_{\mu}\rangle$ and $\langle x_{\mu}\rangle$

$$\begin{cases}\langle p_{\mu}\rangle = \langle\varphi|\boldsymbol{p}_{\mu}|\varphi\rangle = \eta_{\mu\nu}\langle\varphi|\boldsymbol{p}^{\boldsymbol{\nu}}|\varphi\rangle = \eta_{\mu\nu}\langle p^{\nu}\rangle \\ \langle x_{\mu}\rangle = \langle\varphi|\boldsymbol{x}_{\mu}|\varphi\rangle = \eta_{\mu\nu}\langle\varphi|\boldsymbol{x}^{\boldsymbol{\nu}}|\varphi\rangle = \eta_{\mu\nu}\langle x^{\nu}\rangle\end{cases} \tag{9}$$

- the statistical variances-covariances of the operators $\boldsymbol{p}_{\mu}$ and $\boldsymbol{x}_{\boldsymbol{\nu}}$ associated with the state $|\varphi\rangle$ denoted respectively $\langle p_{\mu\nu}\rangle$, $\langle x_{\mu\nu}\rangle$ and $\langle\varrho_{\mu\nu}\rangle$

$$\begin{cases}\langle p_{\mu\nu}\rangle = \langle\varphi|(\boldsymbol{p}_{\boldsymbol{\mu}} - \langle p_{\mu}\rangle)(\boldsymbol{p}_{\boldsymbol{\nu}} - \langle p_{\nu}\rangle)|\varphi\rangle \\ \langle x_{\mu\nu}\rangle = \langle\varphi|(\boldsymbol{x}_{\boldsymbol{\mu}} - \langle x_{\mu}\rangle)(\boldsymbol{x}_{\boldsymbol{\nu}} - \langle x_{\nu}\rangle)|\varphi\rangle \\ \langle\varrho_{\mu\nu}\rangle = \dfrac{1}{2}[\langle\varphi|(\boldsymbol{p}_{\boldsymbol{\mu}} - \langle p_{\mu}\rangle)(\boldsymbol{x}_{\boldsymbol{\nu}} - \langle x_{\nu}\rangle) + (\boldsymbol{x}_{\boldsymbol{\nu}} - \langle x_{\nu}\rangle)(\boldsymbol{p}_{\boldsymbol{\mu}} - \langle p_{\mu}\rangle)|\varphi\rangle\end{cases} \tag{10}$$

A simplest case in which we can have an explicit realization of the relations (9) and (10), is a particular state $|\varphi_0\rangle$ corresponding to a Gaussian-like wavefunction of the form

$$\varphi_0(x) = \langle x|\varphi_0\rangle = e^{iK}\frac{e^{-\mathcal{B}_{\mu\nu}(x^{\mu}-\langle x^{\mu}\rangle)(x^{\nu}-\langle x^{\nu}\rangle)-i\langle p_{\mu}\rangle x^{\mu}}}{[\,(2\pi)^{N}\left|det[\langle x_{\nu}^{\mu}\rangle]\right|]^{1/4}} \tag{11}$$

in which

- the parameters $\mathcal{B}_{\mu\nu}$ are linked with the statistical coordinate variances-covariances $\langle x_{\mu\nu}\rangle$ and coordinate-momenta covariances $\langle\varrho_{\mu\nu}\rangle$ by the relation

$$\mathcal{B}_{\mu\nu} = \frac{1}{4}(\eta_{\mu\rho} + 2i\langle\varrho_{\mu\rho}\rangle)\langle\tilde{x}_{\nu}^{\rho}\rangle \tag{12}$$

with $\langle\tilde{x}_{\nu}^{\rho}\rangle$ the elements of the inverse of the coordinate variances-covariances matrix with elements $\langle x_{\rho}^{\mu}\rangle = \eta^{\mu\nu}\langle x_{\nu\rho}\rangle$ i.e. we have the relations

$$\langle\tilde{x}_{\nu}^{\rho}\rangle\langle x_{\rho}^{\mu}\rangle = \delta_{\nu}^{\mu} = \begin{cases}1\ if\ \mu = \nu \\ 0\ if\ \mu \neq \nu\end{cases} \Leftrightarrow \langle\tilde{x}_{\nu}^{\rho}\rangle\langle x_{\rho\mu}\rangle = \eta_{\nu\mu} \tag{13}$$

- $|det[\langle x_\nu^\mu \rangle]|$ is the absolute value of the determinant of the coordinate variances-covariances matrix with elements $\langle x_\nu^\mu \rangle$.
- $K$ is a real number that doesn't depend on $x^\mu$ ($e^{iK}$ is an unitary complex number).

It can also be established that we have, between $\langle p_{\mu\nu} \rangle$, $\langle \tilde{x}_{\mu\nu} \rangle = \eta_{\mu\rho} \langle \tilde{x}_\nu^\rho \rangle$ and $\langle \varrho_{\mu\nu} \rangle$ the relation

$$\langle p_{\mu\nu} \rangle = \frac{1}{4} \langle \tilde{x}_{\mu\nu} \rangle + \langle \varrho_{\mu\alpha} \rangle \langle \tilde{x}^{\alpha\beta} \rangle \langle \varrho_{\nu\beta} \rangle \qquad (14)$$

The relation (14) is a relativistic multidimensional generalization of the well-known relation between momentum and coordinate variances which correspond to the saturation of the Heisenberg uncertainty relation i.e. the relation

$$(\Delta p)^2 = \frac{1}{4(\Delta x)^2} \qquad (15)$$

Taking into account the relation (13), (14) can be reduced to (15) in the monodimensional case ($N = 1$) and if the coordinate-momenta covariances are equal to zero ($\langle \varrho_{\mu\nu} \rangle = 0$ ). It follows that the state $|\varphi_0\rangle$ can be considered as a generalization of the ground state of a linear harmonic oscillator. It shares some similarities with what are called coherent states and squeezed states in the literature [17-20]. The function in (11) can also be considered as generalization of the Gaussian-like functions introduced and studied in [34-35]

### 3.2 Simplest free particle state compatible with the LCT symmetry

In the current formulation of quantum field theory, the states which are often considered are the states $|p, s\rangle$ of free particles in which the components $p_\mu$ of $p$ are the eigenvalues of the energy-momenta operators $\boldsymbol{p_\mu}$ and $s$ correspond to the spin states (and other possible internal properties of a particle). This description is based on the Wigner classification build with the systematic study of the unitary irreducible representations of the Poincaré group [36-37]. If we consider the simple particular case of a free uncharged scalar boson, the state is just an eigenstate $|p\rangle$ of the momenta operators $\boldsymbol{p_\mu}$.

$$\boldsymbol{p_\mu} |p\rangle = p_\mu |p\rangle \qquad (16)$$

these states correspond to planes waves i.e. the associated "wavefunctions" are of the form

$$\langle x|p\rangle = A e^{-ip_\mu x^\mu} \qquad (17)$$

in which $|x\rangle$ is an eigenstate of the coordinate operators $\boldsymbol{x^\mu}$ and $A$ is a constant amplitude. The wavefunction in (17) is covariant under the action of a Poincaré transformation i.e. the image of a plane wave under the action of the Poincaré group is also a plane wave. But using the equivalence between the LCTs (3) and integral transforms, as shown in the Appendix, it can be deduced that the plane wave form (17) is, in general, not covariant under the action of LCTs (the LCT image of a plane wave is not a plane wave) i.e. we may say that it is not compatible with the LCT symmetry. Contrariwise, it can be verified that the LCT-transform of the Gaussian-like function (11) can be putted in the form

$$\varphi_0(x') = \langle x'|\varphi_0\rangle = e^{iK'} \frac{e^{-\mathcal{B}'_{\mu\nu}(x'^{\mu}-\langle x'^{\mu}\rangle)(x'^{\nu}-\langle x'^{\nu}\rangle)-i\langle p'_{\mu}\rangle x'^{\mu}}}{[\,(2\pi)^N\left|det[\langle x'^{\mu}_{\nu}\rangle]\right|]^{1/4}} \tag{18}$$

in which $e^{iK'}$ is a unitary complex number that doesn't depend on the $x'^{\mu}$. It follows that the state $|\varphi_0\rangle$ corresponding to (11) and (18) may be considered as the simplest state compatible with the LCT symmetry.

Let us denote $\langle p\rangle, \langle x\rangle$ and $\mathcal{P}, \mathcal{X}, \varrho$ the $1 \times N$ and $N \times N$ matrices associated respectively to the means values $\langle p_\mu\rangle, \langle x_\mu\rangle$ and variances-covariances $\langle p_{\mu\nu}\rangle, \langle x_{\mu\nu}\rangle$ and $\langle \varrho_{\mu\nu}\rangle$ of the energy-momenta and coordinates operators $\boldsymbol{p_\mu}$ and $\boldsymbol{x_\mu}$ for the first observer and $\langle p'\rangle, \langle x'\rangle, \mathcal{P}', \mathcal{X}', \varrho'$ their analogs for a second observer. Then, the law of transformations of these parameters corresponding to the LCT transforming the wavefunction (11) into the wavefunction (18) can be written in the form

$$(\langle p'\rangle \quad \langle x'\rangle) = (\langle p\rangle \quad \langle x\rangle)\begin{pmatrix} \mathbb{a} & \mathbb{c} \\ \mathbb{b} & \mathbb{d} \end{pmatrix} \tag{19}$$

$$\begin{pmatrix} \mathcal{P}' & \varrho' \\ \varrho'^T & \mathcal{X}' \end{pmatrix} = \begin{pmatrix} \mathbb{a} & \mathbb{c} \\ \mathbb{b} & \mathbb{d} \end{pmatrix}^T \begin{pmatrix} \mathcal{P} & \varrho \\ \varrho^T & \mathcal{X} \end{pmatrix} \begin{pmatrix} \mathbb{a} & \mathbb{c} \\ \mathbb{b} & \mathbb{d} \end{pmatrix} \tag{20}$$

Now, let us introduce the operators denoted $\boldsymbol{z_\mu}$ and $\boldsymbol{z'_\mu}$ defined by the following relations

$$\begin{cases} \boldsymbol{z_\mu} = \boldsymbol{p_\mu} + 2i\mathcal{B}_{\mu\nu}\boldsymbol{x^\nu} = \boldsymbol{p_\mu} - \langle \varrho_{\mu\nu}\rangle\langle \tilde{x}^{\nu\varrho}\rangle\boldsymbol{x}_\rho + \dfrac{1}{2}i\langle \tilde{x}^{\nu}_{\mu}\rangle\boldsymbol{x}_\nu \\ \boldsymbol{z'_\mu} = \boldsymbol{p'_\mu} + 2i\mathcal{B}'_{\mu\nu}\boldsymbol{x^{\nu\prime}} = \boldsymbol{p'_\mu} - \varrho'_{\mu\nu}\langle \tilde{x}^{\nu\varrho\prime}\rangle\boldsymbol{x'_\rho} + \dfrac{1}{2}i\langle \tilde{x}^{\nu\prime}_{\mu}\rangle\boldsymbol{x'_\nu} \end{cases} \tag{21}$$

It can be established from the relations (11), (12), (13), (18) and (21) that a state $|\varphi_0\rangle$ is a common eigenstate of the operators $\boldsymbol{z_\mu}$ and $\boldsymbol{z'_\mu}$ . Explicitly, the eigenvalues equations are

$$\begin{cases} \boldsymbol{z_\mu}|\varphi_0\rangle = (\langle p_\mu\rangle + 2i\mathcal{B}_{\mu\nu}\langle x^\nu\rangle)|\varphi_0\rangle = \langle z_\mu\rangle|\varphi_0\rangle \\ \boldsymbol{z'_\mu}|\varphi_0\rangle = (\langle p'_\mu\rangle + 2i\mathcal{B}'_{\mu\nu}\langle x^{\nu\prime}\rangle)|\varphi_0\rangle = \langle z'_\mu\rangle|\varphi_0\rangle \end{cases} \quad \forall\, \mu \tag{22}$$

It follows from (22) that we may use the notation $|\langle z\rangle\rangle$ for the state $|\varphi_0\rangle$ i.e.

$$|\langle z\rangle\rangle = |\varphi_0\rangle \tag{23}$$

The eigenvalue equations in (22) can be written as

$$\begin{cases} \boldsymbol{z_\mu}|\langle z\rangle\rangle = \langle z_\mu\rangle|\langle z\rangle\rangle \\ \boldsymbol{z'_\mu}|\langle z\rangle\rangle = \langle z'_\mu\rangle|\langle z\rangle\rangle \end{cases} \tag{24}$$

**Remarks**

- A state $|\langle z\rangle\rangle$ can be considered as a generalization of the state $|p\rangle$ in (16). In fact, we have the following limits

$$\begin{cases}\langle p_{\mu\nu}\rangle \longrightarrow 0 \\ \langle x_{\mu\nu}\rangle \;\rightarrow \infty \\ \langle \varrho^{\times}_{\mu\nu}\rangle \longrightarrow 0\end{cases} \quad \big(\langle \tilde{x}_{\mu\nu}\rangle \longrightarrow 0\big) \Rightarrow \mathcal{B}_{\mu\nu} \rightarrow 0 \Rightarrow \begin{cases}\boldsymbol{z_\mu} = \boldsymbol{p_\mu} + 2i\mathcal{B}_{\mu\nu}\boldsymbol{x^\nu} \longrightarrow \boldsymbol{p_\mu} \\ |\langle z\rangle\rangle \longrightarrow |p\rangle\end{cases} \qquad (25)$$

From a physical point of view, on one side a state $|p\rangle$ describes a case in which the values $p_\mu$ of the momenta of a particle are exactly known and we have the limits in (25) for the variances-covariances. On the other side a state $|\langle z\rangle\rangle$ describes a case in which only the mean values $\langle p_\mu\rangle$ and $\langle x_\mu\rangle$ of the particle coordinates and momenta are known with the variances-covariances $\langle p_{\mu\nu}\rangle, \langle x_{\mu\nu}\rangle$ and $\langle \varrho^{\times}_{\mu\nu}\rangle$ (which have finite values): this case is closer to the physical reality in which there are always quantum fluctuation even in the vacuum (in agreement with the uncertainty principle).

- The function $\langle x|\varphi_0\rangle = \langle x|\langle z\rangle\rangle$ in (11) is square-integrable and then easily normalized.

Like in the Wigner classification [36-37], for a particle having non-zero charges and spin, a state $|\langle z\rangle\rangle$ is to be replaced by a state $|\langle z\rangle, s\rangle$ in which $s$ refer to the spin and charges i.e. internal parameters. In the next section, we consider a study on a spinorial representation of LCTs which show, in particular, that a link can be established between the LCT symmetry and some internal properties of the elementary fermions.

## 4- Spinorial representation of LCTs and properties of elementary fermions

### 4.1 Pseudo-orthogonal representations of LCTs and invariant quadratic operator

In the current formulation of quantum field theory [4], fundamental fermions are described with spinor fields associated to the spinorial representation of the Lorentz group. It follows that in order to formulate a field theory of fermions associated with the LCT symmetry we may look for a spinorial representation of this group. As a spin group is naturally defined to be the double cover of a special orthogonal group, in order to obtain the spinorial representation of LCTs, we begin from the construction of an orthogonal representation. It can be achieved with the introduction of a set of operators called reduced coordinate and momenta operators. In fact, it will be shown that the action of the LCT group on the set of these reduced operators defines a pseudo-orthogonal representation of this group. We introduce the following parameters, operators and notations:

- The parameters denoted $a^\nu_\mu, \mathscr{b}^\nu_\mu$ and $c^\nu_\mu$ which are defined from the variances-covariances $\langle x^\mu_\nu\rangle$, and $\langle \varrho_{\mu\nu}\rangle$ corresponding to a state $|\langle z\rangle\rangle$

$$\begin{cases} a^\mu_\rho a^\rho_\nu = \langle x^\mu_\nu\rangle \\ \mathscr{b}^\nu_\mu = \dfrac{1}{2} a^\rho_\mu \langle \tilde{x}^\nu_\rho\rangle \Leftrightarrow \mathscr{b}^\mu_\rho a^\rho_\nu = \dfrac{1}{2}\delta^\mu_\nu \\ c^\nu_\mu = a^\rho_\mu \langle \varrho_{\rho\lambda}\rangle\langle \tilde{x}^{\lambda\nu}\rangle \end{cases} \qquad (26)$$

The relations (26) means that $a^\nu_\mu, \mathscr{b}^\nu_\mu$ and $c^\nu_\mu$ may be considered as some kind of coordinate-momenta statistical standard deviations corresponding to a state $|\langle z\rangle\rangle$. If we denote $a, \mathscr{b}, c, \mathcal{P}, \mathcal{X}, \varrho$ and $\eta$ the $N \times N$ matrices corresponding respectively to the parameters $a^\nu_\mu, \mathscr{b}^\nu_\mu, c^\nu_\mu, \langle p_{\mu\nu}\rangle, \langle x_{\mu\nu}\rangle, \langle \varrho_{\mu\nu}\rangle$ and $\eta_{\mu\nu}$, it can be established, taking into account (13) and (14) that the relations in (26) is equivalent to the following matrix relations

$$a\mathscr{b} = \mathscr{b}a = \frac{1}{2} \quad \text{(1 being here the } N \times N \text{ identity matrix)} \tag{27}$$

$$\begin{pmatrix} \mathcal{P} & \varrho \\ \varrho^T & \mathcal{X} \end{pmatrix} = \begin{pmatrix} \mathscr{b} & 0 \\ 2ac\mathscr{b} & a \end{pmatrix}^T \begin{pmatrix} \eta & 0 \\ 0 & \eta \end{pmatrix} \begin{pmatrix} \mathscr{b} & 0 \\ 2ac\mathscr{b} & a \end{pmatrix} \tag{28}$$

- The reduced momenta and coordinate operators denoted $\boldsymbol{\mathscr{p}}_{\boldsymbol{\mu}}$ and $\boldsymbol{\mathscr{x}}_{\boldsymbol{\mu}}$

$$\begin{cases} \boldsymbol{\mathscr{p}}_{\boldsymbol{\mu}} = \sqrt{2}a_\mu^\nu(\boldsymbol{p}_{\boldsymbol{\nu}} - \langle p_{\boldsymbol{\nu}} \rangle) - \sqrt{2}c_\mu^\nu(\boldsymbol{x}_\nu - \langle x_{\boldsymbol{\nu}} \rangle) \\ \boldsymbol{\mathscr{x}}_{\boldsymbol{\mu}} = \sqrt{2}\mathscr{b}_\mu^\nu(\boldsymbol{x}_{\boldsymbol{\nu}} - \langle x_{\boldsymbol{\nu}} \rangle) \end{cases} \tag{29}$$

The relation (29) can be considered as a generalization of the construction of a standard normal (Gaussian) variables from ordinary normal random variables. It can be verified that these reduced coordinate and momenta operators satisfy also the canonical commutation relations

$$[\boldsymbol{\mathscr{p}}_\mu, \boldsymbol{\mathscr{x}}_{\boldsymbol{\nu}}] = i\eta_{\mu\nu} \quad [\boldsymbol{\mathscr{p}}_\mu, \boldsymbol{\mathscr{p}}_{\boldsymbol{\nu}}] = 0 \quad [\boldsymbol{\mathscr{x}}_\mu, \boldsymbol{\mathscr{x}}_{\boldsymbol{\nu}}] = 0 \tag{30}$$

However it is to be noticed that in general, for a metric with signature $(N_+, N_-)$, the parameters $a_\rho^\mu, \mathscr{b}_\mu^\nu$ and $c_\mu^\nu$ in (26) and (29) may take complex values if one has $N_- \neq 0$ so the operators $\boldsymbol{\mathscr{p}}_{\boldsymbol{\mu}}$ and $\boldsymbol{\mathscr{x}}_{\boldsymbol{\mu}}$ in (30) are in general not hermitian. If we denote, $\boldsymbol{\mathscr{p}}, \boldsymbol{\mathscr{x}}, \boldsymbol{p}, \boldsymbol{x}, \langle p \rangle$ and $\langle x \rangle$ the $1 \times N$ matrices corresponding respectively to the operators $\boldsymbol{\mathscr{p}}_\mu, \boldsymbol{\mathscr{x}}_\mu, \boldsymbol{p}_\mu, \boldsymbol{x}_\mu$, and the mean values $\langle p_{\boldsymbol{\mu}} \rangle$ and $\langle x_{\boldsymbol{\mu}} \rangle$, the matrix form of the relation (29) is

$$(\boldsymbol{\mathscr{p}} \quad \boldsymbol{\mathscr{x}}) = \sqrt{2}(\boldsymbol{p} - \langle p \rangle \quad \boldsymbol{x} - \langle x \rangle) \begin{pmatrix} a & 0 \\ -c & \mathscr{b} \end{pmatrix} \tag{31}$$

and the inversion of (31) is

$$(\boldsymbol{p} - \langle p \rangle \quad \boldsymbol{x} - \langle x \rangle) = \sqrt{2}(\boldsymbol{\mathscr{p}} \quad \boldsymbol{\mathscr{x}}) \begin{pmatrix} \mathscr{b} & 0 \\ 2ac\mathscr{b} & a \end{pmatrix} \tag{32}$$

we have the following relation (which is a direct consequence of (27))

$$2\begin{pmatrix} \mathscr{b} & 0 \\ 2ac\mathscr{b} & a \end{pmatrix}\begin{pmatrix} a & 0 \\ -c & \mathscr{b} \end{pmatrix} = 2\begin{pmatrix} a & 0 \\ -c & \mathscr{b} \end{pmatrix}\begin{pmatrix} \mathscr{b} & 0 \\ 2ac\mathscr{b} & a \end{pmatrix} = \begin{pmatrix} 1 & 0 \\ 0 & 1 \end{pmatrix} \tag{33}$$

The $N \times N$ matrices $a, \mathscr{b}, c$ satisfy also the following relations $a^T = \eta a \eta, \mathscr{b}^T = \eta \mathscr{b} \eta$ and $c^T = 2\eta a c \mathscr{b} \eta$.
Now let us consider the LCT in (3). It follows from (3) and (19) that we have the following relation

$$(\boldsymbol{p}' - \langle p' \rangle \quad \boldsymbol{x}' - \langle x' \rangle) = (\boldsymbol{p} - \langle p \rangle \quad \boldsymbol{x} - \langle x \rangle) \begin{pmatrix} \mathbb{a} & \mathbb{c} \\ \mathbb{b} & \mathbb{d} \end{pmatrix} \tag{34}$$

Now,we may write the law of transformation of the reduced operators in the form

$$\begin{cases} \boldsymbol{\mathscr{p}}'_\mu = \Pi_\mu^\nu \boldsymbol{\mathscr{p}}_\nu + \Theta_{\boldsymbol{\mu}}^{\boldsymbol{\nu}} \boldsymbol{\mathscr{x}}_\nu \\ \boldsymbol{\mathscr{x}}'_\mu = \Xi_\mu^\nu \boldsymbol{\mathscr{p}}_\nu + \Lambda_{\boldsymbol{\mu}}^{\boldsymbol{\nu}} \boldsymbol{\mathscr{x}}_\nu \end{cases} \Leftrightarrow (\boldsymbol{\mathscr{p}}' \quad \boldsymbol{\mathscr{x}}') = (\boldsymbol{\mathscr{p}} \quad \boldsymbol{\mathscr{x}}) \begin{pmatrix} \Pi & \Xi \\ \Theta & \Lambda \end{pmatrix} \tag{35}$$

Using the relations (31), (32), (33) and their analogs for the second observer, it can be deduced from (35) that

$$(\boldsymbol{p}' - \langle p' \rangle \quad \boldsymbol{x}' - \langle x' \rangle) = 2(\boldsymbol{p} - \langle p \rangle \quad \boldsymbol{x} - \langle x \rangle)\begin{pmatrix} a & 0 \\ -c & \mathscr{b} \end{pmatrix}\begin{pmatrix} \Pi & \Xi \\ \Theta & \Lambda \end{pmatrix}\begin{pmatrix} \mathscr{b}' & 0 \\ 2a'c'\mathscr{b}' & a' \end{pmatrix} \qquad (36)$$

then identifying (36) with (34) and using (5), (28), (30), (33) and their analogs for the second observer, it can be established that we have the following relations

$$\begin{pmatrix} a & 0 \\ -c & \mathscr{b} \end{pmatrix}\begin{pmatrix} \Pi & \Xi \\ \Theta & \Lambda \end{pmatrix} = \begin{pmatrix} \mathbb{a} & \mathbb{c} \\ \mathbb{b} & \mathbb{d} \end{pmatrix}\begin{pmatrix} a' & 0 \\ -c' & \mathscr{b}' \end{pmatrix} \qquad (37)$$

$$\begin{cases} \begin{pmatrix} \Pi & \Xi \\ \Theta & \Lambda \end{pmatrix}^T \begin{pmatrix} \eta & 0 \\ 0 & \eta \end{pmatrix}\begin{pmatrix} \Pi & \Xi \\ \Theta & \Lambda \end{pmatrix} = \begin{pmatrix} \eta & 0 \\ 0 & \eta \end{pmatrix} \\ \begin{pmatrix} \Pi & \Xi \\ \Theta & \Lambda \end{pmatrix}^T \begin{pmatrix} 0 & \eta \\ -\eta & 0 \end{pmatrix}\begin{pmatrix} \Pi & \Xi \\ \Theta & \Lambda \end{pmatrix} = \begin{pmatrix} 0 & \eta \\ -\eta & 0 \end{pmatrix} \end{cases} \Leftrightarrow \begin{cases} \begin{pmatrix} \Pi & \Xi \\ \Theta & \Lambda \end{pmatrix} = \begin{pmatrix} \Pi & -\Theta \\ \Theta & \Pi \end{pmatrix} \\ \Pi^T\eta\Pi + \Theta^T\eta\Theta = \eta \\ \Pi^T\eta\Theta - \Theta^T\eta\Pi = 0 \end{cases} \qquad (38)$$

The $2N \times 2N$ matrix $\begin{pmatrix} \Pi & \Xi \\ \Theta & \Lambda \end{pmatrix} = \begin{pmatrix} \Pi & -\Theta \\ \Theta & \Pi \end{pmatrix}$ belongs to the groups intersection $Sp(2N_+, 2N_-) \cap O(2N_+, 2N_-)$. And it follows from the relations (35) and (38) that the quadratic operator

$$\not{\beth}^+ = \frac{1}{4}\eta^{\mu\nu}(\not{\boldsymbol{p}}_\mu \not{\boldsymbol{p}}_\nu \; + \; \not{\boldsymbol{x}}_\mu \not{\boldsymbol{x}}_\nu) \qquad (39)$$

is invariant under the action of any LCTs. We may call it the invariant quadratic operator associated to the LCT group (it may be considered as a kind of Casimir operator).
The relations (37),(38) define the pseudo-orthogonal representation of LCTs. The relation (37) can be understood as defining a group homomorphism. In fact, let us denote $\mathbb{T} \cong Sp(2N_+, 2N_-)$ the LCT group and $\mathbb{G} \cong Sp(2N_+, 2N_-) \cap O(2N_+, 2N_-)$ the group of the transformations which act on the reduced operators. Let be $\mathbb{t} = \begin{pmatrix} \mathbb{a} & \mathbb{c} \\ \mathbb{b} & \mathbb{d} \end{pmatrix}$ an element of $\mathbb{T}$ and $\mathbb{g} = \begin{pmatrix} \Pi & -\Theta \\ \Theta & \Pi \end{pmatrix}$ an element of $\mathbb{G}$. It follows from the relations (5), (35) and (38) that we have the following definitions

$$\mathbb{T} = \{\mathbb{t} = \begin{pmatrix} \mathbb{a} & \mathbb{c} \\ \mathbb{b} & \mathbb{d} \end{pmatrix} / \begin{pmatrix} \mathbb{a} & \mathbb{c} \\ \mathbb{b} & \mathbb{d} \end{pmatrix}^T \begin{pmatrix} 0 & \eta \\ -\eta & 0 \end{pmatrix}\begin{pmatrix} \mathbb{a} & \mathbb{c} \\ \mathbb{b} & \mathbb{d} \end{pmatrix} = \begin{pmatrix} 0 & \eta \\ -\eta & 0 \end{pmatrix}\} \qquad (40)$$

$$\mathbb{G} = \{\mathbb{g} = \begin{pmatrix} \Pi & -\Theta \\ \Theta & \Pi \end{pmatrix} / \begin{pmatrix} \Pi & -\Theta \\ \Theta & \Pi \end{pmatrix}^T \begin{pmatrix} \eta & 0 \\ 0 & \eta \end{pmatrix}\begin{pmatrix} \Pi & -\Theta \\ \Theta & \Pi \end{pmatrix} = \begin{pmatrix} \eta & 0 \\ 0 & \eta \end{pmatrix}\} \qquad (41)$$

Then the relation (37) can be considered as defining a homomorphism $\not{f}$ from $\mathbb{T}$ to $\mathbb{G}$

$$\begin{gathered} \not{f}: \mathbb{T} \longrightarrow \mathbb{G} \\ \begin{pmatrix} \mathbb{a} & \mathbb{c} \\ \mathbb{b} & \mathbb{d} \end{pmatrix} \mapsto \begin{pmatrix} \Pi & -\Theta \\ \Theta & \Pi \end{pmatrix} \end{gathered} \qquad (42)$$

The homomorphism $\not{f}$ is not injective so the orthogonal representation of $\mathbb{T}$ defined by it on the set of reduced operators is not faithful. However it is surjective and is defined for any LCT. Now we may introduce the set of elements of $\mathbb{T}$ which leave invariant the reduced operators i.e. the kernel of the homomorphism $\not{f}$ . This set is a normal subgroup of $\mathbb{T}$ that we denote $\mathbb{H}$

$$\mathbb{H} = \{\mathbb{h} \in \mathbb{T}/\not{f}(\mathbb{h}) = \begin{pmatrix} \Pi & -\Theta \\ \Theta & \Pi \end{pmatrix} = \begin{pmatrix} 1 & 0 \\ 0 & 1 \end{pmatrix}\} = Ker(\not{f}) \tag{43}$$

As the homomorphism $\not{f}$ is surjective, we have $Im(\not{f}) = \mathbb{G}$ and it follows from the fundamental homomorphism theorem that we have the isomorphism

$$Im(\not{f}) \cong \mathbb{T}/Ker(\not{f}) \Leftrightarrow \mathbb{G} \cong \mathbb{T}/\mathbb{H} \tag{44}$$

### 4.2 Properties and structure of the group $\mathbb{G} \cong \mathbb{T}/\mathbb{H} \cong Sp(2N_+, 2N_-) \cap O(2N_+, 2N_-)$

It can be shown that the group $\mathbb{G}$ corresponding to the pseudo-orthogonal representation of LCTs is isomorph to the pseudo-unitary group $U(N_+, N_-)$ i.e.

$$\mathbb{G} \cong \mathbb{T}/\mathbb{H} \cong Sp(2N_+, 2N_-) \cap O(2N_+, 2N_-) \cong U(N_+, N_-) \tag{45}$$

In order to prove (45), let us consider the operator

$$\not{\boldsymbol{z}}_\mu = a_\mu^\nu(\boldsymbol{z}_{\boldsymbol{\nu}} - \langle z_\nu \rangle) = \frac{1}{\sqrt{2}}(\not{\boldsymbol{p}}_\mu + i\not{\boldsymbol{x}}_{\boldsymbol{\mu}}) \tag{46}$$

If $\not{\boldsymbol{z}}$ is the $1 \times N$ matrix corresponding to the $\not{\boldsymbol{z}}_\mu$, it can be deduced from (35), (38) and (46) that the law of transformation of $\not{\boldsymbol{z}}$ is

$$\not{\boldsymbol{z}}' = \not{\boldsymbol{z}}(\Pi - i\Theta) \tag{47}$$

If the $N \times N$ matrix $\Omega = \Pi - i\Theta$ is introduced, the following equivalence can be established

$$\Omega^\dagger \eta \Omega = \eta \Leftrightarrow \begin{cases} \Pi^T \eta \Pi + \Theta^T \eta \Theta = \eta \\ \Pi^T \eta \Theta - \Theta^T \eta \Pi = 0 \end{cases} \tag{48}$$

The relation $\Omega^\dagger \eta \Omega = \eta$ means that $\Omega$ is an element of the pseudo-unitary group $U(N_+, N_-)$. Then the comparison of (48) with (38) lead to the isomorphism (45).

Some of the main properties of the group $U(N_+, N_-) \cong \mathbb{G}$ can be found in the existing literature [38-39]. It is for instance known that $U(N_+, N_-)$ is connected and its maximal compact subgroup is $U(N_+) \times U(N_-)$.The dimension of $U(N_+, N_-)$ itself is equal to $N^2 = (N_+ + N_-)^2$ and the dimension of its maximal compact subgroup $U(N_+) \times U(N_-)$ is equal to $(N_+)^2 + (N_-)^2$. It is to be noticed that the main source of non-compactness is the indefinite signature $(N_+, N_-)$. In fact, it is well known that in the case $N_- = 0$ (or $N_+ = 0$), the group $U(N)$ is compact.

It is also well-known that the rank of $U(N)$ (i.e. the topological dimension of its maximal torus) is equal to $N$. It is also the dimension (as vectorial space) of the maximal abelian subalgebra (i.e. the Cartan subalgebra) of its Lie algebra $\mathfrak{u}(N)$. It follows that the rank of the maximal compact subgroup $U(N_+) \times U(N_-)$ of $U(N_+, N_-)$ is also equal to $N_+ + N_- = N$.

For the case of the group $\mathbb{G} \cong U(1,4)$ which leads to the classification given in the table 1 below, it follows that: $\mathbb{G}$ is connected, its dimension is equal to 25, the dimension of its maximal compact subgroup, isomorph to $U(1) \times U(4)$, is equal to 17. The rank of this maximal compact subgroup is equal to 5: the fact that the classification in the table 1 is defined with five commutative operators is directly related to this property.

As the group $U(N_+, N_-)$ in the isomorphism (45) is connected it follows that we must have

$$\mathbb{G} \cong \mathbb{T}/\mathbb{H} \cong Sp(2N_+, 2N_-) \cap O(2N_+, 2N_-) \cong Sp(2N_+, 2N_-) \cap SO_0(2N_+, 2N_-) \quad (49)$$

in which $SO_0(2N_+, 2N_-)$ is the identity component of $SO(2N_+, 2N_-)$. The topological double cover of $SO_0(2N_+, 2N_-)$ is the spin group $Spin(2N_+, 2N_-)$ [40-42]. It is this relation that can be exploited for the construction of the spinorial representation of the group $\mathbb{G}$.

**Remark**: if we consider the law of transformation (47) associated to the operator $\boldsymbol{\mathcal{z}}_\mu$ in (46) and their conjugate, it can be noticed that we obtain transformations which are similar to the Bogolioubov ones [21-22]

### 4.3 Link between the spinorial representation of the LCTs and fermions properties

A spinorial representation of LCTs can be deduced from the pseudo-orthogonal representation defined through the group homomorphism $\mathcal{f}$ in (42). Like this pseudo-orthogonal, this spinorial representation which is a spinorial representation of the group $\mathbb{G}$ is not a faithful representation of $\mathbb{T}$. However, it is defined for any element of $\mathbb{T}$. To obtain explicitly this spinorial representation, we need to introduce the operator

$$\mathbb{p} = \alpha^\mu \boldsymbol{\mathcal{p}}_\mu + \beta^\mu \boldsymbol{\mathcal{x}}_\mu \quad (50)$$

in which $\alpha^\mu$ and $\beta^\mu$ are the generators of the Clifford algebra $\mathcal{C\ell}(2N_+, 2N_-)$. They verify the following anticommutation relations

$$\begin{cases} \alpha^\mu \alpha^\nu + \alpha^\nu \alpha^\mu = 2\eta^{\mu\nu} \\ \beta^\mu \beta^\nu + \beta^\nu \beta^\mu = 2\eta^{\mu\nu} \\ \alpha^\mu \beta^\nu + \beta^\nu \alpha^\mu = 0 \end{cases} \quad (51)$$

Let us denote $\mathbb{S}$ the topological double cover of $\mathbb{G}$, $\mathbb{S}$ is a subgroup of $Spin(2N_+, 2N_-)$. The relation between $\mathbb{S}$ and $\mathbb{G}$ is defined by a covering map $\mathcal{u}$ according to the relation

$$\begin{cases} \mathcal{u}: \mathbb{S} \longrightarrow \mathbb{G} \\ \quad \mathcal{S} \longmapsto \mathbb{g} = \begin{pmatrix} \Pi & -\Theta \\ \Theta & \Pi \end{pmatrix} \end{cases} \Leftrightarrow \begin{cases} (\boldsymbol{\mathcal{p}}' \quad \boldsymbol{\mathcal{x}}') = (\boldsymbol{\mathcal{p}} \quad \boldsymbol{\mathcal{x}}) \begin{pmatrix} \Pi & -\Theta \\ \Theta & \Pi \end{pmatrix} \\ \qquad \mathbb{p}' = \mathcal{S} \mathbb{p} \mathcal{S}^{-1} \end{cases} \quad (52)$$

Given (29), (37) and (42), it follows that (52) define also a spinorial representation of $\mathbb{T}$.Like the pseudo-orthogonal representation (42), this representation is defined for any LCTs but it is not faithful. From a physical point of view, this “unfaithfulness” correspond to the fact that some types of LCT do not affect the “internal states” of particles.

Any element $\Sigma$ of the Lie algebra $\mathfrak{spin}(2N_+, 2N_-)$ of $Spin(2N_+, 2N_-)$ can be written in the form [40-42]

$$\Sigma = \xi_{\mu\nu} \alpha^\mu \alpha^\nu + \vartheta_{\mu\nu} \beta^\mu \beta^\nu + \theta_{\mu\nu} \alpha^\mu \beta^\nu \quad (53)$$

in which the components $\xi_{\mu\nu}$ and $\vartheta_{\mu\nu}$ satisfy the constraints

$$\begin{cases} \xi_{\mu\nu} = -\xi_{\nu\mu} \\ \vartheta_{\mu\nu} = -\vartheta_{\nu\mu} \end{cases} \quad (54)$$

and there is no constraint on the $\theta_{\mu\nu}$. The relations (53) and (54) correspond to the fact that the topological dimension of $Spin(2N_+, 2N_-)$ is equal to $N(2N-1)$. But, given the constraints (38) on the elements of $\mathbb{G}$, an element of the Lie algebra $\mathfrak{s} \subset \mathfrak{spin}(2N_+, 2N_-)$ of the double cover $\mathbb{S}$ of $\mathbb{G}$ must also satisfy the following additional constraint

$$\begin{cases} \xi_{\mu\nu} = \vartheta_{\mu\nu} \\ \theta_{\mu\nu} = \theta_{\nu\mu} \end{cases} \tag{55}$$

It follows that the dimension of $\mathfrak{s}$ is equal to $N^2$. It is also as expected the topological dimension of $\mathbb{S}$ and $\mathbb{G}$. It follows from (53), (54) and (55) that we can choose as a basis of $\mathfrak{s}$ the family

$$\mathfrak{B} = \{\frac{1}{2}(\alpha^\mu\alpha^\nu + \beta^\mu\beta^\nu), \frac{1}{2}(\alpha^\mu\beta^\nu + \alpha^\nu\beta^\mu)\} \tag{56}$$

For the case $(N_+, N_-) = (1,4)$, the dimension of $\mathfrak{s}$ is equal to $N^2 = 25$. Some of the infinitesimal generators of $\mathbb{S}$ given in (56) corresponds to its non-compact part and some of them correspond to its compact part (i.e. its maximal compact subgroup):

- The infinitesimal generators which correspond to the non-compact part are those which corresponds to the "hyperbolic rotations" i.e. the family

$$\mathfrak{B}_{nc} = \{\frac{1}{2}(\alpha^0\alpha^j + \beta^0\beta^j), \frac{1}{2}(\alpha^0\beta^j + \alpha^j\beta^0)\} \quad \text{for } j = 1,2,3,4 \tag{57}$$

- The infinitesimal generators which correspond to the compact part are those which corresponds to the "ordinary circular rotations" i.e. the family

$$\mathfrak{B}_c = \{\frac{1}{2}(\alpha^j\alpha^l + \beta^j\beta^l), \alpha^\mu\beta^\mu, \frac{1}{2}(\alpha^j\beta^l + \alpha^j\beta^l)\}\ \} \quad \text{for } j,l = 1,2,3,4 \tag{58}$$

the number of elements of $\mathfrak{B}_{nc}$ is equal to 8 and the number of the elements of $\mathfrak{B}_c$ is equal to 17: these numbers are respectively, as expected, equal to the dimensions of the non-compact part of $\mathbb{S}$ (or of $\mathbb{G}$) and the dimension of its maximal compact subgroup. The dimension of the Cartan subalgebra of the algebra generated by the set $\mathfrak{B}_c$ is equal to 5 (which is also the rank of the maximal compact subgroup of $\mathbb{S}$ or of $\mathbb{G}$). The following five commutative hermitian generators can be chosen to be a basis of this Cartan subalgebra

$$\mathcal{Y}^0 = \frac{1}{2}i\alpha^0\beta^0 \quad \mathcal{Y}^1 = \frac{1}{3}i\alpha^1\beta^1 \quad \mathcal{Y}^2 = \frac{1}{3}i\alpha^2\beta^2 \quad \mathcal{Y}^3 = \frac{1}{3}i\alpha^3\beta^3 \quad \mathcal{Y}^4 = \frac{1}{2}i\alpha^4\beta^4 \tag{59}$$

A classification of a generation of fermions of the Standard model, can be deduced from this spinorial representation corresponding to a space with signature (1,4). It can be achieved explicitly if we choose exactly for the generators $\alpha^\mu$ and $\beta^\mu$ of the Clifford algebra $\mathcal{C\ell}(2,8)$ the following matrix representations

$$\begin{cases} \alpha^0 = \sigma^1 \otimes \sigma^0 \otimes \sigma^0 \otimes \sigma^0 \otimes \sigma^0 & \beta^0 = \sigma^2 \otimes \sigma^0 \otimes \sigma^0 \otimes \sigma^0 \otimes \sigma^0 \\ \alpha^1 = i\sigma^3 \otimes \sigma^1 \otimes \sigma^0 \otimes \sigma^0 \otimes \sigma^0 & \beta^1 = -i\sigma^3 \otimes \sigma^2 \otimes \sigma^0 \otimes \sigma^0 \otimes \sigma^0 \\ \alpha^2 = i\sigma^3 \otimes \sigma^3 \otimes \sigma^1 \otimes \sigma^0 \otimes \sigma^0 & \beta^2 = -i\sigma^3 \otimes \sigma^3 \otimes \sigma^2 \otimes \sigma^0 \otimes \sigma^0 \\ \alpha^3 = i\sigma^3 \otimes \sigma^3 \otimes \sigma^3 \otimes \sigma^1 \otimes \sigma^0 & \beta^3 = -i\sigma^3 \otimes \sigma^3 \otimes \sigma^3 \otimes \sigma^2 \otimes \sigma^0 \\ \alpha^4 = i\sigma^3 \otimes \sigma^3 \otimes \sigma^3 \otimes \sigma^3 \otimes \sigma^1 & \beta^4 = -i\sigma^3 \otimes \sigma^3 \otimes \sigma^3 \otimes \sigma^3 \otimes \sigma^2 \end{cases} \tag{60}$$

,with $\sigma^0 = \begin{pmatrix} 1 & 0 \\ 0 & 1 \end{pmatrix}$ $\sigma^1 = \begin{pmatrix} 0 & 1 \\ 1 & 0 \end{pmatrix}$ $\sigma^2 = \begin{pmatrix} 0 & -i \\ i & 0 \end{pmatrix}$ $\sigma^3 = \begin{pmatrix} 1 & 0 \\ 0 & -1 \end{pmatrix}$, define the operators

$$\begin{cases} I_3 = \frac{1}{2}\mathcal{Y}^0 - \frac{1}{2}\mathcal{Y}^4 \quad Y_W = \mathcal{Y}^0 + \mathcal{Y}^1 + \mathcal{Y}^2 + \mathcal{Y}^3 + \mathcal{Y}^4 \\ Q = \mathcal{Y}^0 + \frac{1}{2}\mathcal{Y}^1 + \frac{1}{2}\mathcal{Y}^2 + \frac{1}{2}\mathcal{Y}^3 = I_3 + \frac{Y_W}{2} \end{cases} \tag{61}$$

and then identify the eigenvalues of $I_3, Y_W$ and $Q$ respectively with the weak isospin, weak hypercharges and electric charges of a generation of elementary fermions. The obtained classification is given in the table 1 below.

It is worth pointing out that the operators $\mathcal{Y}^0, \mathcal{Y}^1, \mathcal{Y}^2, \mathcal{Y}^3$ and $\mathcal{Y}^4$ which describe the “internal states” of the particles as shown in the table 1 are infinitesimal generators of the maximal compact subgroup the group $\mathbb{S}$.

**Table 1:** *Classification of leptons, quarks, (and their antiparticles) belonging to a family according to the eigenvalues of the operators* $\mathcal{Y}^0, \mathcal{Y}^1, \mathcal{Y}^2, \mathcal{Y}^3$, $\mathcal{Y}^4$, $I_3$, $Y_W$ *and* $Q$

| N° | $\mathcal{Y}^0$ | $\mathcal{Y}^1$ | $\mathcal{Y}^2$ | $\mathcal{Y}^3$ | $\mathcal{Y}^4$ | $I_3$ | $Y_W$ | $Q$ | Particle |
|---|---|---|---|---|---|---|---|---|---|
| 1 | -1/2 | -1/3 | -1/3 | -1/3 | -1/2 | 0 | -2 | -1 | $e_R$ |
| 2 | -1/2 | -1/3 | -1/3 | -1/3 | 1/2 | -1/2 | -1 | -1 | $e_L$ |
| 3 | -1/2 | -1/3 | -1/3 | 1/3 | -1/2 | 0 | -4/3 | -2/3 | $\bar{u}_R^{blue}$ |
| 4 | -1/2 | -1/3 | -1/3 | 1/3 | 1/2 | -1/2 | -1/3 | -2/3 | $\bar{u}_L^{blue}$ |
| 5 | -1/2 | -1/3 | 1/3 | -1/3 | -1/2 | 0 | -4/3 | -2/3 | $\bar{u}_R^{green}$ |
| 6 | -1/2 | -1/3 | 1/3 | -1/3 | 1/2 | -1/2 | -1/3 | -2/3 | $\bar{u}_L^{green}$ |
| 7 | -1/2 | -1/3 | 1/3 | 1/3 | -1/2 | 0 | -2/3 | -1/3 | $d_R^{red}$ |
| 8 | -1/2 | -1/3 | 1/3 | 1/3 | 1/2 | -1/2 | 1/3 | -1/3 | $d_L^{red}$ |
| 9 | -1/2 | 1/3 | -1/3 | -1/3 | -1/2 | 0 | -4/3 | -2/3 | $\bar{u}_R^{red}$ |
| 10 | -1/2 | 1/3 | -1/3 | -1/3 | 1/2 | -1/2 | -1/3 | -2/3 | $\bar{u}_L^{red}$ |
| 11 | -1/2 | 1/3 | -1/3 | 1/3 | -1/2 | 0 | -2/3 | -1/3 | $d_R^{green}$ |
| 12 | -1/2 | 1/3 | -1/3 | 1/3 | 1/2 | -1/2 | 1/3 | -1/3 | $d_L^{green}$ |
| 13 | -1/2 | 1/3 | 1/3 | -1/3 | -1/2 | 0 | -2/3 | -1/3 | $d_R^{blue}$ |
| 14 | -1/2 | 1/3 | 1/3 | -1/3 | 1/2 | -1/2 | 1/3 | -1/3 | $d_L^{blue}$ |
| 15 | -1/2 | 1/3 | 1/3 | 1/3 | -1/2 | 0 | 0 | 0 | $\bar{\nu}_R$ |
| 16 | -1/2 | 1/3 | 1/3 | 1/3 | 1/2 | -1/2 | 1 | 0 | $\bar{\nu}_L$ |
| 17 | 1/2 | -1/3 | -1/3 | -1/3 | -1/2 | 1/2 | -1 | 0 | $\nu_L$ |
| 18 | 1/2 | -1/3 | -1/3 | -1/3 | 1/2 | 0 | 0 | 0 | $\nu_R$ |
| 19 | 1/2 | -1/3 | -1/3 | 1/3 | -1/2 | 1/2 | -1/3 | 1/3 | $\bar{d}_L^{blue}$ |
| 20 | 1/2 | -1/3 | -1/3 | 1/3 | 1/2 | 0 | 2/3 | 1/3 | $\bar{d}_R^{blue}$ |
| 21 | 1/2 | -1/3 | 1/3 | -1/3 | -1/2 | 1/2 | -1/3 | 1/3 | $\bar{d}_L^{gree}$ |
| 22 | 1/2 | -1/3 | 1/3 | -1/3 | 1/2 | 0 | 2/3 | 1/3 | $\bar{d}_R^{green}$ |
| 23 | 1/2 | -1/3 | 1/3 | 1/3 | -1/2 | 1/2 | 1/3 | 2/3 | $u_L^{red}$ |
| 24 | 1/2 | -1/3 | 1/3 | 1/3 | 1/2 | 0 | -4/3 | 2/3 | $u_R^{red}$ |
| 25 | 1/2 | 1/3 | -1/3 | -1/3 | -1/2 | 1/2 | 1/3 | 1/3 | $\bar{d}_L^{red}$ |
| 26 | 1/2 | 1/3 | -1/3 | -1/3 | 1/2 | 0 | 2/3 | 1/3 | $\bar{d}_R^{red}$ |
| 27 | 1/2 | 1/3 | -1/3 | 1/3 | -1/2 | 1/2 | 1/3 | 2/3 | $u_L^{green}$ |
| 28 | 1/2 | 1/3 | -1/3 | 1/3 | 1/2 | 0 | -4/3 | 2/3 | $u_R^{green}$ |
| 29 | 1/2 | 1/3 | 1/3 | -1/3 | -1/2 | 1/2 | 1/3 | 2/3 | $u_L^{blue}$ |
| 30 | 1/2 | 1/3 | 1/3 | -1/3 | 1/2 | 0 | -4/3 | 2/3 | $u_R^{blue}$ |
| 31 | 1/2 | 1/3 | 1/3 | 1/3 | -1/2 | 1/2 | 1 | 1 | $\overline{e_L}$ |
| 32 | 1/2 | 1/3 | 1/3 | 1/3 | 1/2 | 0 | 2 | 1 | $\overline{e_R}$ |

The table 1 corresponds to a generation of fermions. The example of the first generation may be considered: $e_L$ is the left-handed electron and $\overline{e_L}$ is its antiparticle. $e_R$ is the right-handed electron and $\overline{e_R}$ its antiparticle, $\nu_L$ and $\overline{\nu_L}$ are respectively the left-handed neutrino and its antiparticle. The existence of right-handed sterile neutrino, $\nu_R$, is suggested by this table 1. The up and down quarks are respectively denoted $u$ and $d$ and their antiparticles $\bar{u}$ and $\bar{d}$. The lower script refers to chirality (R for right-handed and L for left-handed) and the upper script are colors (blue, green or red). The existence of the three possible quarks colors is described by the different combinations of the eigenvalues of the operators $\mathcal{Y}^1, \mathcal{Y}^2$ and $\mathcal{Y}^3$.

**Remark:** It can be deduced from the relations (39), (50) and (51) that we have between the invariant quadratic operator $\beth^+$ in (39) and the operator $\mathbb{p}$ in (50) the relation

$$(\mathbb{p})^2 = 4\beth^+ + i\eta_{\mu\nu}\alpha^\mu\beta^\nu \Leftrightarrow \beth^+ = \frac{1}{4}[(\mathbb{p})^2 - i\eta_{\mu\nu}\alpha^\mu\beta^\nu] \qquad (62)$$

### 4.4 Description of a generation of fermions with a single field and unified theories

The table 1 suggests the description of a generation of fermions of the standard model with a single spinor field $\overrightarrow{\boldsymbol{\psi}}$ . $\overrightarrow{\boldsymbol{\psi}}$ belong to a spinor space $\mathfrak{F}$ and can be written in the form

$$\overrightarrow{\boldsymbol{\psi}} = \boldsymbol{\psi}^{\boldsymbol{a}}\vec{\zeta}_a \qquad (63)$$

in which the $\vec{\zeta}_a$ are the $2^5 = 32$ common eigenspinors of the operators $\mathcal{Y}^0, \mathcal{Y}^1, \mathcal{Y}^2, \mathcal{Y}^3, \mathcal{Y}^4$ $(a = 1,2, \dots ,32)$. $\{\vec{\zeta}_a\}$ is a canonical basis of $\mathfrak{F}$. The element $\mathcal{S} \in \mathbb{S}$ defined from the relation (52) acts on an element $\overrightarrow{\boldsymbol{\psi}}$ of $\mathfrak{F}$

$$\overrightarrow{\boldsymbol{\psi}}' = \mathcal{S}\overrightarrow{\boldsymbol{\psi}} \Leftrightarrow \boldsymbol{\psi}'^{\boldsymbol{a}} = \mathcal{S}^a_{b}\boldsymbol{\psi}^{\boldsymbol{b}} \qquad (64)$$

$\overrightarrow{\boldsymbol{\psi}}'$ is the image of $\overrightarrow{\boldsymbol{\psi}}$ under the action of an LCT corresponding to $\mathcal{S}$ through the relations (37), (42) and (52). The $\mathcal{S}^a_{b}$ are the element of the matrix of $\mathcal{S}$ in the basis $\{\vec{\zeta}_a\}$ of $\mathfrak{F}$.
From the relations (50), (52) and (64), it can be deduced that an LCT-covariant equation which may correspond to the spinor field $\overrightarrow{\boldsymbol{\psi}}$ in (63) is of the form

$$\mathbb{p}\overrightarrow{\boldsymbol{\psi}} = \epsilon\overrightarrow{\boldsymbol{\psi}} \Leftrightarrow (\alpha^\mu\not{\boldsymbol{p}}_\mu + \beta^\mu\not{\boldsymbol{x}}_\mu)\overrightarrow{\boldsymbol{\psi}} = \epsilon\overrightarrow{\boldsymbol{\psi}} \qquad (65)$$

We may perform a comparison between the above description and the formulations that are used in the framework of the Standard model [43-46] and some grand unified theories [43].
Let us consider, to begin, the case of the formulation of Quantum Chromodynamics based on the $SU(3)_C$ group. In this formulation, an element $U$ of $SU(3)$ acts on the field

$$\boldsymbol{\psi} = \begin{pmatrix} \boldsymbol{\psi_r} \\ \boldsymbol{\psi_g} \\ \boldsymbol{\psi_b} \end{pmatrix} \qquad (66)$$

In which each of the color components $\boldsymbol{\psi_r}, \boldsymbol{\psi_g}$ and $\boldsymbol{\psi_b}$ are Dirac spinors with 4 components associated to the action of the group $Spin$ (1,3) (i.e. spinorial representation of the Lorentz group). So, if we take into account spacetime symmetry, the field $\boldsymbol{\psi}$ in (66) has $3 \times 4 = 12$ components. In this formulation of QCD, the “spacetime symmetry” and the “internal color symmetry” are combined in a trivial way (in agreement with the Coleman-Mandula no-go theorem [47-48]). In other word (if we also take into account spacetime translation symmetry), the full symmetry group associated to the field in (65) is the direct product (trivial combination) of the Poincaré group and $SU(3)_C$.
Now if we consider the full Standard model [43-46], it is known that the symmetry group associated to the internal symmetries is $SU(3)_C \times SU(2)_W \times U(1)_Y$ . The gauge group $SU(2)_W \times U(1)_Y$ corresponds to the electroweak theory in which weak and electromagnetism interactions are unified but there is no unification between them and strong interaction which correspond to $SU(3)_C$ .
In the grand unified theories i.e. GUTs [43], the unification of electroweak and strong interactions is expected to happen at some very high energy scale and another larger gauge group is considered instead of the Standard model one. However, like the Standard model, these GUTs do not consider a non-trivial combination of spacetime and internal symmetries. In other words, they don’t try to circumvent the Coleman-Mandula theorem. It is known that there are unification models that go beyond and circumvent this no-go theorem [49-50]. However, to our knowledge, the introduction of the LCT group which corresponds to a “spacetime-energy-momentum symmetry” (i.e. phase space symmetry) in these contexts of unifications is considered for the first time in the present work.

In a generation of fermions of the Standard model, we have 2 leptons and 2 quarks (each quark can have one of the 3 different colors). If we take into account this color multiplicity, a generation of the standard model should be described with $2 + (2 \times 3) = 8$ spinor fields (particle, antiparticle and their possible chiral states are described with one spinor field). If the existence of right handed neutrinos is supposed, each of these spinor field is equivalent to a Dirac spinor which has 4 components. So the total number of components is

$$(2 \times 4) + (2 \times 3 \times 4\ ) = 32 \tag{67}$$

It follows that we may describe a generation of the Standard model with a field having 32 components. But as remarked previously, in the case of the Standard model and GUTs, the symmetry group corresponding to the field is a trivial combination of the spacetime symmetry group (a non-compact group) and the internal symmetry group (a compact group).

Now the relation (63) suggests also the description of a generation of fermions with a single field having 32 components. But unlike the previous cases, there is in this description a non-trivial combination of “phase space symmetry” with what are called “internal symmetries”. Moreover, what are called “internal symmetries” can just be identified as corresponding to the compact part of the “phase space symmetry” i.e. associated, as it was seen, to the maximal compact subgroup of the group $\mathbb{S}$. It is a natural and non-trivial combination that may lead to an unified theory of fundamental interactions. We may also remark that the Coleman-Mandula theorem does not apply in this case because we did not consider only “spacetime symmetry” but the symmetry corresponding to the LCT group which is a “phase space symmetry”.

The results obtained through this section suggests that it may be possible to build a unified theory of interactions using the LCT group as a gauge group. As this group describes a mixing between spacetime coordinate operators and momenta-energy operators, it is expected that gravitation could be included naturally in this kind of unification. In the next section, it is shown that the main groups that are currently considered in the gauge theories of gravitation [51-53] can be obtained from the contraction of the LCT group.

## 5- Obtaining the group $O(N_+, N_-)$ from the contraction of an LCT group

### 5.1 General description

The purpose of this section is to highlight the fact that it is possible to obtain the indefinite orthogonal group $O(N_+, N_-)$ from the contraction of the LCT group $Sp(2N_+, 2N_-)$.

This contraction may be approached with the Inönü-Wigner concept of group contraction which is for instance considered in the references [5, 54-60]. Well-known examples of applications of this concept in physics are the obtaining of the Galilei group from the contraction of the Poincare group if the speed of light tends to infinity $c \to \infty$ [54, 56] and the obtaining of the Poincare group as contraction of the de Sitter group if the de Sitter radius tends to infinity or equivalently if the cosmological constant $\Lambda$ associated to this radius tends to zero: $\Lambda \to 0$ [5].

However, our goal is not to perform a complete and systematic study of the contraction of $Sp(2N_+, 2N_-)$, in the framework of this Inönü-Wigner concept, but just to highlight the fact that $O(N_+, N_-)$ can be obtained "among" the results of this kind of contraction when quantum and gravitational effects are neglected (see remarks below).

The obtained results are applied in the particular cases of a quadridimensional theory and a pentadimensional theory. For convenience, the SI units is used through this section. To highlight the possibility of a contraction of the LCT group $Sp(2N_+, 2N_-)$ which leads to the indefinite orthogonal group $O(N_+, N_-)$, two parameters $\wp$ and $\ell$ may be introduced so that instead of (3), we write the following parameterization for an LCT

$$\begin{cases} \boldsymbol{p}'_\mu = \mathbb{a}_\mu^\nu \boldsymbol{p}_\nu + \dfrac{(\wp)^2}{\hbar} \mathbb{b}_{\boldsymbol{\mu}}^{\boldsymbol{\nu}} \boldsymbol{x}_\nu \\ \boldsymbol{x}'_\mu = \dfrac{(\ell)^2}{\hbar} \mathbb{c}_\mu^\nu \boldsymbol{p}_\nu + \mathbb{d}_{\boldsymbol{\mu}}^{\boldsymbol{\nu}} \boldsymbol{x}_\nu \end{cases} \Longrightarrow \begin{pmatrix} \mathbb{a} & \dfrac{(\ell)^2}{\hbar} \mathbb{c} \\ \dfrac{(\wp)^2}{\hbar} \mathbb{b} & \mathbb{d} \end{pmatrix} = e^{\begin{pmatrix} \lambda + \frac{\ell \wp}{\hbar} \mu & \frac{(\ell)^2}{\hbar} \varphi \\ \frac{(\wp)^2}{\hbar} \vartheta & \lambda - \frac{\ell \wp}{\hbar} \mu \end{pmatrix}} \tag{68}$$

The exponential form in (68) means that the $2N \times 2N$ matrix $\begin{pmatrix} \lambda + \frac{\ell \wp}{\hbar} \mu & \frac{(\ell)^2}{\hbar} \varphi \\ \frac{(\wp)^2}{\hbar} \vartheta & \lambda - \frac{\ell \wp}{\hbar} \mu \end{pmatrix}$ should belong to the Lie algebra $\mathfrak{sp}(2N_+, 2N_-)$ of $Sp(2N_+, 2N_-)$. We have to remark here as it is known that the exponential map is in general not surjective for the case of a symplectic group. However it is also known that any elements of the group can be obtained from the composition of elements that can be putted in the exponential form. So we may consider the exponential form (68) through our discussion without losing generality. The relations in (68) defines effectively an LCT if and only if we have the following relations for the $N \times N$ matrices $\mathbb{a}, \mathbb{b}, \mathbb{c}, \mathbb{d}, \lambda, \mu, \vartheta$ and $\varphi$

$$\begin{cases} \mathbb{a}^T \eta \mathbb{d} - \dfrac{(\wp)^2 (\ell)^2}{(\hbar)^2} \mathbb{b}^T \eta \mathbb{c} = \eta \\ \mathbb{a}^T \eta \mathbb{b} - \mathbb{b}^T \eta \mathbb{a} = 0 \\ \mathbb{c}^T \eta \mathbb{d} - \mathbb{d}^T \eta \mathbb{c} = 0 \end{cases} \Longrightarrow \begin{cases} \lambda^T = -\eta \lambda \eta & \mu^T = \eta \mu \eta \\ \vartheta^T = \eta \vartheta \eta & \varphi^T = \eta \varphi \eta \end{cases} \tag{69}$$

Unlike in the parameterization corresponding to (3), the coefficients of the matrices $\mathbb{a}, \mathbb{b}, \mathbb{c}, \mathbb{d}, \lambda, \varphi, \vartheta$ and $\mu$ involved in (68) and (69) are all dimensionless if it is supposed that $\wp$ has the dimension of momentum and $\ell$ the dimension of length. This fact may be considered as an advantage of (68) and (69) compared to (3). Another main utility of the parameterization

(68) is its importance in the study of the possibility of obtaining the indefinite orthogonal group $O(N_{+},N_{-})$ from the contraction of the LCT group $Sp(2N_{+},2N_{-})$.

If the possibility of obtaining the Lorentz group $O(1,3)$ from the contraction of the LCT group $Sp(2,6)$ is for instance considered, it is expected that it occurs when some "quantum effects" related to the Heisenberg uncertainty relations are neglected. As these uncertainty relations relate the coordinates and momenta statistical standard deviations (i.e. uncertainties), we may propose the following hypothesis: the values of the parameters $\wp$ and $\ell$ are respectively equal or proportional (very close) to the minimum possible values of the momentum and coordinates standard deviations. In the state of the current knowledge in astrophysics and particle physics, two fundamental constants may be respectively related to these minimum possible values: the first one is the cosmological constant $\Lambda$ [5-9] and the second one is the Planck length $\ell_P$ [9, 61]. The cosmological constant is often considered as related to the vacuum fluctuation so it may be considered as natural to relate it with the minimum values of momenta standard deviations. The Planck length is usually considered as a kind of minimum length so it may be related to the minimum values of coordinates standard deviations. Explicitly, we may suppose that the current values of $\wp$ and $\ell$ are respectively (in SI units).

$$\begin{cases} \wp = \hbar\sqrt{\Lambda} \simeq 1{,}083.10^{-60} kg.m.s^{-1} \\ \ell = \ell_P = \sqrt{\dfrac{\hbar G}{c^3}} \simeq 1{,}616..10^{-35} m \end{cases} \tag{70}$$

in which $\hbar$ is the reduced Planck constant, $\Lambda$ is the cosmological constant, $\ell_P$ is the Planck length, $G$ is the gravitational constant and $c$ is the speed of light. Taking into account (70), the relation (68) becomes

$$\begin{cases} \boldsymbol{p}'_{\mu} = \mathbb{a}_{\mu}^{\nu}\boldsymbol{p}_{\nu} + (\hbar\Lambda)\mathbb{b}_{\boldsymbol{\mu}}^{\boldsymbol{\nu}}\boldsymbol{x}_{\nu} \\ \boldsymbol{x}'_{\mu} = (\dfrac{G}{c^3})\mathbb{c}_{\mu}^{\nu}\boldsymbol{p}_{\nu} + \mathbb{d}_{\boldsymbol{\mu}}^{\boldsymbol{\nu}}\boldsymbol{x}_{\nu} \end{cases} \Rightarrow \begin{pmatrix} \mathbb{a} & (\dfrac{G}{c^3})\mathbb{c} \\ (\hbar\Lambda)\mathbb{b} & \mathbb{d} \end{pmatrix} = e^{\begin{pmatrix} \lambda+\sqrt{\frac{\hbar\Lambda G}{c^3}}\mu & (\frac{G}{c^3})\varphi \\ (\hbar\Lambda)\vartheta & \lambda-\sqrt{\frac{\hbar\Lambda G}{c^3}}\mu \end{pmatrix}} \tag{71}$$

It can be deduced from the relation (71) that the contraction of the LCT group $Sp(2N_{+},2N_{-})$ which leads to the group $O(N_{+},N_{-})$ occurs in the limits $\hbar\Lambda \to 0$ and $\frac{G}{c^3} \to 0$. In these limits, the relation (71) becomes

$$\begin{cases} \boldsymbol{p}'_{\mu} = \mathbb{a}_{\mu}^{\nu}\boldsymbol{p}_{\nu} \\ \boldsymbol{x}'_{\mu} = \mathbb{d}_{\boldsymbol{\mu}}^{\boldsymbol{\nu}}\boldsymbol{x}_{\nu} \end{cases} \Rightarrow \begin{pmatrix} \mathbb{a} & 0 \\ 0 & \mathbb{d} \end{pmatrix} = e^{\begin{pmatrix} \lambda & 0 \\ 0 & \lambda \end{pmatrix}} \Leftrightarrow \mathbb{a} = \mathbb{d} = e^{\lambda} \tag{72}$$

According to the relation (69), the $N \times N$ matrix $\lambda$ fulfills the relations $\lambda^T = -\eta\lambda\eta$ . It means that $\lambda$ belongs to the Lie algebra $\mathfrak{o}(N_{+},N_{-})$ of the Lie group $O(N_{+},N_{-})$ and it follows that the $N \times N$ matrix $\mathbb{a} = \mathbb{d} = e^{\lambda}$ itself belongs to $O(N_{+},N_{-})$. The indefinite orthogonal group $O(N_{+},N_{-})$ can be obtained, as expected, from the contraction of the LCT group $Sp(2N_{+},2N_{-})$.

**Remarks**

- The limits $\hbar\Lambda \to 0$ and $\frac{G}{c^3} \to 0$ corresponding to the contraction of the LCT group $Sp(2N_{+},2N_{-})$ as described above is to be understood, more exactly, as corresponding to the limits $\hbar \to 0$ and $G \to 0$ ($\Lambda$ keeps a non-zero value and $c$ still has a finite value). From a physical point of view, these limits means that quantum and gravitational effects are neglected. For the signature (1,4), the limits $\Lambda \to 0$ and $c \to +\infty$ corresponds respectively to the contractions of the de Sitter group which leads to the Poincare group and the contraction of the Poincare group which leads to the Galilei Group [5, 54, 56, 59].

- Rigorously, we must understand the "obtaining" of the indefinite group $O(N_+, N_-)$ from the contraction of the group $Sp(2N_+, 2N_-)$, as highlighted above, as just showing the fact that $O(N_+, N_-)$ can be considered as a "part" of the results of this contraction if it is understood in the exact framework of the Inönü-Wigner concept (like the homogeneous Lorentz group can be considered as a "part" of the results of the contraction of de Sitter group when $\Lambda \to 0$ while the "full result" of this contraction is the Poincare group i.e. the inhomogeneous Lorentz group). A complete and systematic study of this contraction may bring more interesting results.

### 5.2 Quadridimensional Theory

In the case of a quadridimensional theory corresponding to a space with signature (1,3), the LCT group is the pseudo-symplectic group $Sp\ (2,6)$. The group contraction, described through the relation (71) and (72), corresponding to the limits $\hbar \to 0$ and $G \to 0$ then corresponds to the contraction of this group $Sp\ (2,6)$ which leads to the Lorentz group $O(1,3)$. It is compatible with the fact that the Lorentz group corresponds to the theory of special relativity in which quantum and gravitational effects are neglected.

**Remark**: instead of the LCT corresponding to the relations (71) and (72), an LCT coupled with a translation (an Affine Canonical Transformation) may be considered

$$\begin{cases} \boldsymbol{p}'_\mu = \mathbb{a}_\mu^\nu \boldsymbol{p}_\nu + \hbar\Lambda \mathbb{b}_{\boldsymbol{\mu}}^{\boldsymbol{\nu}} \boldsymbol{x}_\nu + \hbar\Lambda \mathbb{e}_{\boldsymbol{\mu}} \\ \boldsymbol{x}'_\mu = \dfrac{G}{c^3} \mathbb{c}_\mu^\nu \boldsymbol{p}_\nu + \mathbb{d}_{\boldsymbol{\mu}}^{\boldsymbol{\nu}} \boldsymbol{x}_\nu + \mathbb{f}_{\boldsymbol{\mu}} \end{cases} \tag{73}$$

Then in the limits $\hbar \to 0$ , $G \to 0$ and $\Lambda \to 0$, the contraction of the group corresponding to the transformation (73) leads to the Poincaré group.

### 5.3 Pentadimensional theory

The results described in the section 4 concerning the relation between the spinorial representation of LCTs and elementary fermions suggest that an LCT group corresponding to a pentadimensional space with signature (1, 4) is an interesting case to be considered. For this pentadimensional case, the LCT group is the pseudo-symplectic group $Sp(2,8)$. In the limits $\hbar \to 0$ and $G \to 0$ as described through the relations (71) and (72), the contraction of this group $Sp(2,8)$ leads to the indefinite orthogonal group $O(1,4)$. It means that the de Sitter group, which corresponds to the de Sitter invariant Special Relativity [5-9], can be obtained from this contraction. If this contraction is followed by another contraction corresponding to the limit $\Lambda \to 0$, the Poincare group is obtained [5, 59].

The relation between the spinorial representation of LCTs and properties of elementary fermions of the Standard Model highlighted in the section 4 and the group contraction relations described above suggest that the LCT group $Sp(2,8)$ may be used to formulate a unified theory of fundamental interactions which include naturally a quantum theory of gravity. A gauge theory based on this LCT group may be, for instance, considered for this purpose.

## 6- Main results

Some of the main results established through this works are the followings:

***i)*** The group formed by the LCTs defined through the relations (3), (4) and (5) may be considered as a symmetry group for a Relativistic Quantum Theory. Main covariant relations corresponding to this group are the canonical commutations relations.

***ii)*** In the most general multidimensional case, the LCTs can be considered as the elements of a pseudo-symplectic group $Sp(2N_+, 2N_-)$which acts on the set of energy-momenta and coordinates operators corresponding to a $N$-dimensional pseudo-Euclidian space with signature $(N_+, N_-)$. $N = N_+ + N_-$

***iii)***There is an equivalence between the integral transforms, known in the framework of signal processing and optics, and the linear operators' transformations called LCTs in the framework of relativistic quantum physics as considered in this work (see appendix). There is also a similarity between them and Bogolioubov transformation.

***iv)*** Lorentz transformations and fractional Fourier transforms are particular cases of LCTs.

***v)*** The introduction of the reduced momenta and coordinates operators defined in the relation (29) leads to a pseudo-orthogonal and a spinorial representation of the LCT group. In the framework of a pentadimensional theory with signature (1,4) , this spinorial representation can be used to explain some properties of a generation of fermions and to classify them. This classification suggests the existence of sterile neutrinos and the possibility of describing a fermions generation with a single field. Sterile neutrinos are among the most interesting research topics that are currently underway [62-64]

***vi)*** The Lorentz group can be obtained from the contraction of the LCT group $Sp(2,6)$ in the limits $\hbar \to 0$ and $G \to 0$. The de Sitter group and the Poincaré group can be obtained from the contractions of the LCT group $Sp(2,8)$ respectively in the limits $(\hbar, G) \to (0,0)$ and $(\hbar, G, \Lambda) \to (0,0,0)$ . Intuitively, the limits $(\hbar, G) \to (0,0)$ means that quantum and gravitational effects are neglected.

***vii)*** It may be possible to establish a unified theory of fundamental interactions, which includes naturally a quantum theory of gravity, using an LCT group as gauge group.

## 7- Discussions, conclusions and perspectives

The definition of LCTs as linear transformations combining coordinates and momenta operators and keeping invariant the canonical commutation relations provides a simple and natural way to perform multidimensional generalization. It makes easier their study within the framework of relativistic quantum physics in which they may be considered as the elements of a symmetry group.

LCTs can be considered at the same time as generalization of fractional Fourier transforms (which are themselves generalizations of Fourier transforms) and as generalization of Lorentz transformations.

On one hand, Fourier transforms is deeply related to quantum physics through the wave - particle duality. They are mainly present in the description of the link between the coordinates and momenta representations. Coordinates representation may be seen as highlighting the corpuscle nature of a particle while momentum representation highlights its wave nature (According to the well-known Planck-Einstein-De Broglie relations, energy and momentum are equivalent to the frequency and wave vector). In a certain point of view, the fractional Fourier and more generally LCTs permit to approach the wave-particle duality in a more subtle way.

On the other hand it is well known that Lorentz group is the main symmetry group of the current formulation of many relativistic theories. It is then natural to find that an LCTs group, which includes Fourier and Lorentz Transformations, may be expected to be a symmetry group of relativistic quantum physics. It is seen, through this work, that this idea is quantitatively supported by the fact that LCTs are the simplest transformations which keep invariants the canonical commutations relations. Our approach gives an adequate quantitative formulation of a covariance principle for relativistic quantum physics.

As already discussed at the ends of the sections 4 and 5, the relations that can be established between LCTs and the elementary fermions of the Standard Model through the spinorial representation described in the section 4 and the results obtained from the group contraction study in the section 5 suggest that the LCT group corresponding to a pentadimensional theory may be used in the establishment of an unified theory of fundamental interactions.

There are other possible applications of the obtained results that can be considered by analyzing the relations (11), (12), (14), (21), (26) and (35). It can be noticed in these relations that the statistical variances–covariance of coordinates and momenta are at the core of the formulation that is considered. But it is well known, from the kinetic theory of gases for instance, that thermodynamic variables are linked with the statistical variance of particles speeds and then with the statistical variance of their momenta. So it may be expected that the introduction of the statistical variances–covariances that is considered here can also be understood within a quantum thermodynamic framework. This idea can also be supported explicitly by the fact the wavefunction in (11) which is a generalization of the Gaussian-like function introduced in [34-35] can be used to establish an LCT-covariant phase space representation of quantum theory. This phase space representation can provide phase space distributions analogous to the one that Wigner introduced in an attempt to introduce quantum corrections in thermodynamic [65]. That is to say that the results that are obtained here may be also exploited for the formulation of a relativistic quantum thermodynamic compatible with the LCT symmetry. This may in particular help to answer some interesting basic questions like the one considered in [66].

The presence of the statistical variance-covariance of coordinates and momenta which introduce a Gaussian factor in the wavefunctions (11), (18) of particles can also be exploited to deal with the divergence problem in quantum field theory. It follows that the theory that can be developed with an LCT group can be naturally a divergence free theory. This characteristic can also be considered as fundamentally linked with the fact that minimum possible values of momenta and coordinates standard deviations can be associated with the LCT symmetry as discussed in the section 5.

In brief, it may be expected that the formulation of a relativistic quantum field theory based on the LCT symmetry can provide a natural way for the establishment of a divergence free unified theory of fundamental interactions, including a quantum theory of gravity, and which is coupled with a relativistic quantum thermodynamic. This kind of theory would be of great use for theoretical physics, astrophysics and cosmology. However, further studies are needed to assess the validity of these perspectives.

**Acknowledgements:** The authors are grateful to the INSTN-Madagascar and its staff for their help during the conception and completion of this work and to the reviewers for their stimulating remarks.

**Appendix: Equivalence of integral transforms and linear operators' transformations**

In the framework of signal processing and optics, Linear Canonical Transformations (LCTs) are well-known to be integral transforms which admit as particular case some classical useful integral transformations like Fourier and fractional Fourier ones. Let $\psi$ be a function of a variable $t$ (usually considered as the time variable). A linear canonical transformation which transforms $\psi$ to a function $\Psi$ of a variable $t'$ can be defined by the following relation

$$\Psi(t') = C\int \psi(t) e^{\frac{i}{\mathbb{c}}\left(t't - \frac{\mathbb{a}(t')^2 + \mathbb{d}(t)^2}{2}\right)} dt \qquad (A.1)$$

with $\mathbb{a}$, $\mathbb{c}$, and $\mathbb{d}$ elements of a $2 \times 2$ matrix $\mathfrak{M} = \begin{pmatrix} \mathbb{a} & \mathbb{c} \\ \mathbb{b} & \mathbb{d} \end{pmatrix}$ belonging to the special linear group $SL(2)$ i.e. $det\mathfrak{M} = \mathbb{a}\mathbb{d} - \mathbb{b}\mathbb{c} = 1$ and $C$ a complex number which may depend on $\mathbb{a}, \mathbb{b}, \mathbb{c}$ and $\mathbb{d}$.

For the case $\mathbb{a} = \mathbb{d} = cos\vartheta$ and $\mathbb{c} = -\mathbb{b} = -sin\vartheta$, the transformation (A.1) is a fractional Fourier transform and for the case $\vartheta = \frac{\pi}{2}$, it is a Fourier transform.

If two linear operators $\boldsymbol{t}$ and $\boldsymbol{\omega}$, associated respectively to time and angular frequency and, satisfying the canonical commutation relation $[\boldsymbol{\omega}, \boldsymbol{t}]_- = i$ are introduced, it can be shown that the integral transform (A.1) is equivalent to a linear operators transformation which keeps invariant the canonical commutation relation

$$\begin{cases} \boldsymbol{\omega'} = \mathbb{a}\boldsymbol{\omega} + \mathbb{c}\boldsymbol{t} \\ \boldsymbol{t'} = \mathbb{b}\boldsymbol{\omega} + \mathbb{d}\boldsymbol{t} \\ [\boldsymbol{\omega'}, \boldsymbol{t'}]_- = [\boldsymbol{\omega}, \boldsymbol{t}]_- = i \end{cases} \qquad (A.2)$$

The last relation (invariance of canonical commutation relation) leads directly to the relation $\mathbb{a}\mathbb{d} - \mathbb{b}\mathbb{c} = 1$. It is straightforward to remark that (A.2) is a particular case of the LCTs defined through the relations (3) and (4) for the dimension $N = 1$ and with the identification $\boldsymbol{\omega} = \boldsymbol{p^0}, \boldsymbol{t} = \boldsymbol{x^0}$ and $\eta_{00} = 1$.

To prove the equivalence between the integral transforms (A.1) and the operators' transformations (A.2), let us consider the following formalism: let $|t\rangle$ be an eigenvector (eigenstate) of the operator $\boldsymbol{t}$ and $\psi(t)$ the $\boldsymbol{t}$-representation of a state vector $|\psi\rangle$ i.e. $\psi(t) = \langle t|\psi\rangle$. The following (quantum theory like) relations are assumed

$$\begin{cases} \langle t|\boldsymbol{t}|\psi\rangle = t\langle t|\psi\rangle = t\psi(t) \\ \langle t|\boldsymbol{\omega}|\psi\rangle = i\frac{\partial}{\partial t}\langle t|\psi\rangle = i\frac{\partial}{\partial t}\psi(t) \end{cases} \Leftrightarrow \begin{cases} \langle \psi|\boldsymbol{t}|t\rangle = t\langle \psi|t\rangle = t\psi^*(t) \\ \langle \psi|\boldsymbol{\omega}|t\rangle = (\langle t|\boldsymbol{\omega}|\psi\rangle)^* = -i\frac{\partial}{\partial t}\psi^*(t) \end{cases} \qquad (A.3)$$

Similar relations stand for the pair $(\boldsymbol{\omega'}, \boldsymbol{t'})$. Then, the transformation (A.1) may be interpreted as a change from the $t$- representation to the $t'-$ representation

$$\Psi(t') = \langle t'|\psi\rangle = \int \langle t'|t\rangle\langle t|\psi\rangle dt = \int \psi(t)\langle t'|t\rangle dt \qquad (A.4)$$

Then, to prove the equivalence between (A.1) and (A.2), it is just needed to prove that

$$\langle t'|t\rangle = Ce^{\frac{i}{\mathbb{c}}\left(t't - \frac{\mathbb{a}(t')^2 + \mathbb{d}(t)^2}{2}\right)} \qquad (A.5)$$

The combination of the transformation in (A.2) and the relations (A.3) gives

$$\begin{cases}\langle t'|\boldsymbol{\omega}'|t\rangle = \mathbb{a}\langle t'|\boldsymbol{\omega}|t\rangle + \mathbb{b}\langle t'|\boldsymbol{t}|t\rangle \\ \langle t'|\boldsymbol{t}'|t\rangle = \mathbb{c}\langle t'|\boldsymbol{\omega}|t\rangle + \mathbb{d}\langle t'|\boldsymbol{t}|t\rangle\end{cases} \Leftrightarrow \begin{cases} i\dfrac{\partial\langle t'|t\rangle}{\partial t'} = \mathbb{a}(-i\dfrac{\partial\langle t'|t\rangle}{\partial t}) + \mathbb{b}t\langle t'|t\rangle \\ t'\langle t'|t\rangle = \mathbb{c}(-i\dfrac{\partial\langle t'|t\rangle}{\partial t}) + \mathbb{d}t\langle t'|t\rangle\end{cases} \qquad (A.6)$$

The resolution of the differential equations system in (A.6) (taking into account the relation ($\mathbb{a}\mathbb{d} - \mathbb{c}\mathbb{b} = 1$) leads to the relation (A.5). The LCTs satisfying the relations (3) and (4) in the sect. 2 are the multidimensional generalization of (A.2) and then equivalently generalization of (A.1) too.